\documentclass[prd,twocolumn,tightenlines,superscriptaddress,nofootinbib,showpacs]{revtex4}
\usepackage{amssymb,latexsym}
\usepackage{amsmath,amsbsy,bbm}
\usepackage{epsfig,bm}
\usepackage{graphicx,comment}
\usepackage{color}
\begin{document} 

\title{Subtlety of Determining the Critical Exponent $\nu$ of the Spin-1/2
Heisenberg Model with a Spatially Staggered Anisotropy on the Honeycomb 
Lattice}

\author{F.-J.~Jiang}
\email[]{fjjiang@itp.unibe.ch}
\affiliation{Center for Research and Education in Fundamental Physics, 
Institute for Theoretical Physics, Bern University, 
Sidlerstrasse 5, CH-3012 Bern, Switzerland}

\author{U.~Gerber}
\email[]{gerberu@itp.unibe.ch}
\affiliation{Center for Research and Education in Fundamental Physics, 
Institute for Theoretical Physics, Bern University, 
Sidlerstrasse 5, CH-3012 Bern, Switzerland}

\vspace{-1cm}
  
\begin{abstract}

Puzzled by the indication of a new critical theory for the  
spin-1/2 Heisenberg model with a spatially staggered anisotropy on the 
square lattice as suggested in \cite{Wenzel08}, we study a similar 
anisotropic spin-1/2 Heisenberg model on the honeycomb lattice. 
The critical point where the phase transition occurs due to the 
dimerization as well as the critical exponent 
$\nu$ are analyzed in great detail. Remarkly, using most of the available
data points in conjunction with the expected finite-size scaling ansatz with 
a sub-leading correction indeed leads to a consistent $\nu = 0.691(2)$ with that calculated 
in \cite{Wenzel08}. However by using the data with large number of spins $N$, 
we obtain $\nu = 0.707(6)$ which agrees with the  
most accurate Monte Carlo $O(3)$ value $\nu = 0.7112(5)$ as well.

\end{abstract}
%\pacs{12.39.Fe, 75.10.Jm, 75.40.Mg, 75.50.Ee}

\maketitle

\section{Introduction}
The discovery of high temperature (high $T_c$) superconductivity in the cuprate 
materials has triggered vigorous research investigation on 
spin-1/2 Heisenberg-type models, which are argued and believed to be the 
correct models for the undoped precursors of high $T_c$ cuprates (undoped antiferromagnets).
In addition to analytic results, highly accurate first principles Monte Carlo
studies of Heisenberg-type models are available due to the fact 
that these models on geometrically non-frustrated lattices
do not suffer from a sign problem, which in turn 
allows one to design efficient cluster algorithms to simulate
these systems. Indeed, thanks to the advance in numerical 
algorithms as well as the increasing power of computing
resources, the undoped antiferromagnets are among the quantitatively 
best-understood condensed matter systems. 
For example, the low-energy
parameters of the spin-1/2 Heisenberg model on the square lattice, which are 
obtained from the combination of Monte Carlo calculation and the 
corresponding low-energy effective field theory, are in quantitative
agreement with the experimental results \cite{Wie94}.  

Anisotropic Heisenberg models have been studied intensely
during the last twenty years. Besides their 
phenomenological importance, they are of great interest from a 
theoretical perspective as well \cite{Parola93,Affleck96,Sandvik99,Irkhin00,Kim00}. 
For example, due to the newly discovered pinning effects of 
the electronic liquid 
crystal in the underdoped cuprate
superconductor YBa$_2$Cu$_3$O$_{6.45}$ \cite{Hinkov2007,Hinkov2008}, 
the Heisenberg model 
with spatially anisotropic couplings $J_{1}$ and $J_{2}$ has
attracted theoretical interest \cite{Pardini08,Jiang09}.
Further, 
numerical evidence indicates that the anisotropic
Heisenberg model with staggered arrangement of the antiferromagnetic 
couplings may belong to a new universality class, in contradiction
to the theoretical $O(3)$ universality predictions \cite{Wenzel08}. In particular, 
the numerical Monte Carlo studies of other
anisotropic spin-1/2 Heisenberg models on the square lattice, 
for example, the ladder and the plaquette models,
seem to be described well by the 3-dimensional classical Heisenberg
universality class \cite{Alb08,Wenzel09} even at small volumes. 
These studies indicate that the 
staggered arrangement of the anisotropy in the spin-1/2 Heisenberg model 
might be responsible for the unexpected results found in \cite{Wenzel08}. 
In order to 
clarify this issue further, based on the expectation that 
the honeycomb lattice is one of the candidates to investigate whether such 
a new universality class does exist, 
we have simulated the spin-1/2 Heisenberg model with
spatially staggered anisotropy on the honeycomb lattice.

The motivation of our study is to investigate whether the unconventional critical behavior
found in \cite{Wenzel08} can be observed again by considering the 
Heisenberg model with a similar anisotropic pattern 
on the honeycomb lattice. In particular, one can expect such kind of 
investigation might shed some light on 
understanding the discrepancy between $\nu = 0.689(5)$ determined in
\cite{Wenzel08} and the most accurate Monte Carlo value $\nu=0.7112(5)$ 
for the $O(3)$ universality 
class \cite{Cam02}. To achieve this goal, the critical behavior of the 
spin-1/2 Heisenberg model with a spatially staggered anisotropy 
has been investigated in great detail in this study. In particular, the 
transition point driven by the anisotropy as well as 
the critical exponent $\nu$ are determined with high precision by 
fitting the numerical data points to their predicted critical behavior near 
the transition. We focus on the critical exponent $\nu$ so that we can perform
a more thorough analysis and obtain data with as large number
of spins as possible. Although we find a consistent $\nu = 0.691(2)$ with that
calculated in \cite{Wenzel08} by using most of the available data points as well
as taking into account the correction term in the finite-size scaling ansatz,
we obtain $\nu=0.707(6)$ which agrees with the known $O(3)$ Monte Carlo value
as well when only the data points with a large number of spins are used in
the fits.

This paper is organized as follows. In section \ref{micro}, the anisotropic
Heisenberg model and the relevant observables studied in this work are briefly described. 
Section \ref{results} contains our numerical results.
In particular, the corresponding critical point as well as the critical
exponent $\nu$ are determined by fitting the numerical data to their predicted 
critical behavior near the transition. Finally, we conclude our study 
in section \ref{discussion}.

\begin{figure}
\begin{center}
\includegraphics[width=0.43\textwidth]{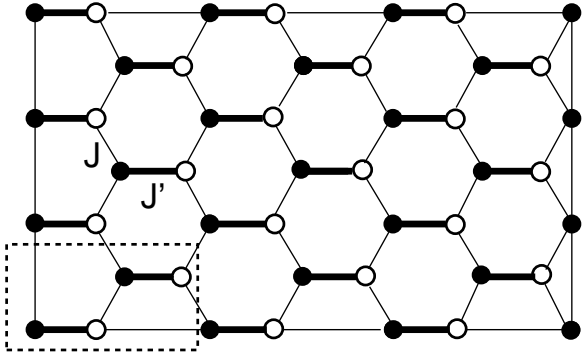}
\end{center}\vskip-0.5cm
\caption{The anisotropic spin-1/2 Heisenberg model investigated in this study.}
\label{fig0}
\end{figure}
     
\section{Microscopic Model and Corresponding Observables}
\label{micro}
In this section we introduce the Hamiltonian of the microscopic Heisenberg 
model as well as some relevant observables. The Heisenberg 
model considered in this study is defined by the Hamilton operator
\begin{eqnarray}
\label{hamilton}
H = \sum_{\langle xy \rangle}J\,\vec S_x \cdot \vec S_{y}
+\sum_{\langle x'y' \rangle}J'\,\vec S_{x'} \cdot \vec S_{y'},
\end{eqnarray}
where $J'$ and $J$ are antiferromagnetic exchange couplings connecting
nearest neighbor spins $\langle  xy \rangle$ 
and $\langle x'y' \rangle$, respectively. Figure 1 illustrates the Heisenberg
model described by eq.~(\ref{hamilton}) on the honeycomb lattice with periodic 
spatial boundary conditions implemented in our simulations. 
The dashed rectangle in figure 1, which contains $4$ spins, is the elementary
cell for building a periodic honeycomb lattice covering a rectangular area.  
For instance, the honeycomb lattice shown in figure 1 contains 3 $\times$ 3 
elementary cells. The honeycomb lattice is not a
Bravais lattice. 
Instead it consists of two triangular Bravais sub-lattices $A$ and $B$
(depicted by solid and open circles in figure 1). As a consequence, 
the momentum space of the honeycomb lattice is a doubly-covered Brillouin 
zone of the two triangular sub-lattices. 
To study the critical behavior of this anisotropic Heisenberg model near 
the transition driven by the dimerization, in particular to determine 
the critical exponent $\nu$, several relevant observables are
measured in our simulations. A physical quantity of central interest is the 
staggered susceptibility 
(corresponding to the third component of the staggered magnetization $M_s^3$) 
which is given by
\begin{eqnarray}
\label{defstagg}
\chi_s &=& \frac{1}{L_{1} L_{2}} 
\int_0^\beta dt \ \langle M^3_s(0) M^3_s(t) \rangle \nonumber \\
&=& \frac{1}{L_{1} L_{2}} \int_0^\beta dt \ \frac{1}{Z} 
\mbox{Tr}[M^3_s(0) M^3_s(t) \exp(- \beta H)].
\end{eqnarray}
Here $\beta$ is the inverse temperature, $L_{1}$ and $L_{2}$ are the spatial 
box sizes in the $1$- and $2$-direction, respectively, and 
$Z = \mbox{Tr}\exp(- \beta H)$ is the partition function. The staggered 
magnetization order parameter 
$\vec{M}_s$ is defined as $\vec M_s = \sum_x (-1)^{x} \vec S_x$. Here
$(-1)^{x} = 1$ on the $A$ sublattice and $(-1)^{x} = -1$ on the $B$ 
sublattice, respectively.
Other relevant quantities calculated from our simulations are the spin stiffnesses in 
the $1$- and $2$-directions
\begin{eqnarray}
\rho_{si} = \frac{1}{\beta L^2_{i}}\langle W^2_{i}\rangle,
\end{eqnarray}
here $i \in \{1,2\}$ refers to the spatial directions and $W^2_{i}$ is
the winding number squared in the $i$-direction.
Finally, the second-order Binder cumulant for the staggered magnetization, 
which is defined as
\begin{eqnarray}
Q_{2} = \frac{\langle (M^3_{s})^4 \rangle}{\langle (M^3_{s})^2 \rangle^2}
\end{eqnarray}
are measured as well. By carefully investigating the spatial volume 
(and possibly the space-time volume) and the $J'/J$ dependence of
these observables, one can determine the transition point as well
as the critical exponent $\nu$ with high precision.

\begin{figure}
\begin{center}
\includegraphics[width=0.43\textwidth]{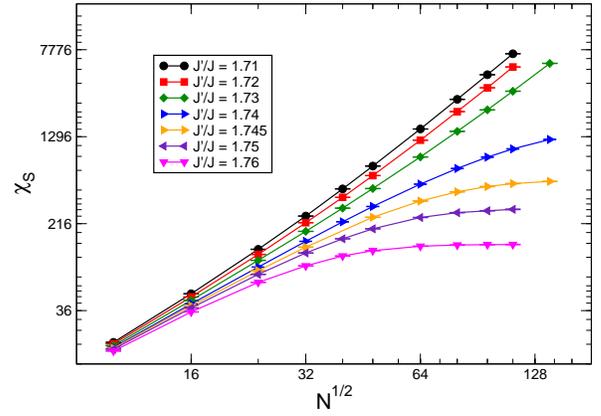}
\end{center}
\vskip-0.5cm
\caption{The space-time volume dependence of the staggered susceptibilities
$\chi_s$ of the Heisenberg model in this work for various values 
of anisotropy $J'/J$. Here $N$ stands for the number of spins. The lines are 
added to guide the eye.}
\label{fig1}
\end{figure}

\begin{figure*}
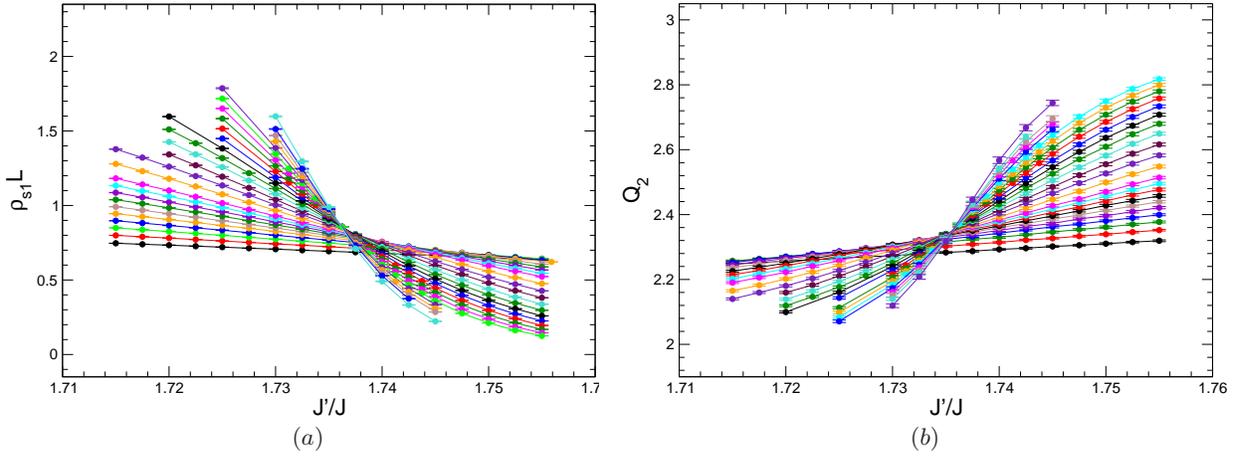

\begin{tabular}{cc}
\includegraphics[width=0.45\textwidth]{rhos_all_1.eps}
&\includegraphics[width=0.45\textwidth]{smd_all_1.eps}
\\
$(a)$ & $(b)$
\end{tabular}
\vskip-0.25cm
\caption{The $J'/J$ dependence of $\rho_{s1}L\,\, (a)$ and $Q_2\,\, (b)$.
The lines are added to guide the eye.}
\label{fig2}
\end{figure*}

\section{Determination of the Critical Point and the 
Critical Exponent $\nu$}
\label{results}

\begin{figure}
\begin{center}
\includegraphics[width=0.43\textwidth]{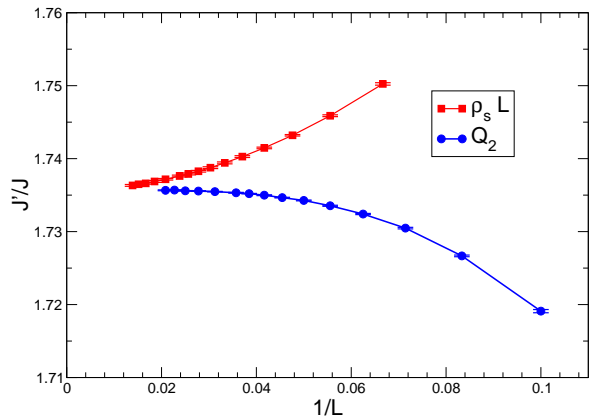}
\end{center}
\vskip-0.5cm
\caption{Crossing points of $\rho_{s1}L$ and $Q_2$ obtained from
the corresponding observables at system sizes $L$ and $2L$. The lines are
added to guide the eye.}
\label{fig3}
\end{figure}

\begin{figure}
\begin{center}
\includegraphics[width=0.415\textwidth]{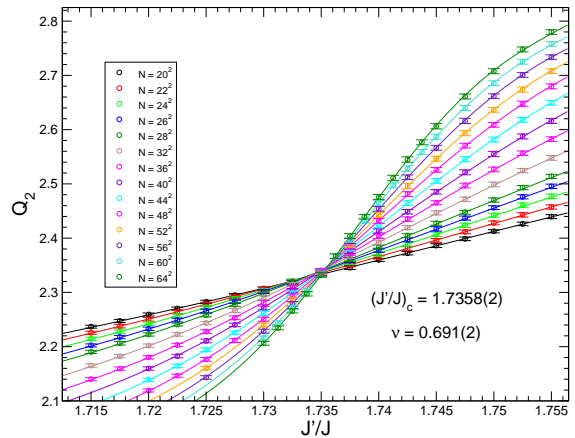}
\end{center}
\vskip-0.5cm
\caption{Fit of most of the numerical data points for $Q_2$ to their finite-size scaling ansatz
including the correction term $cL^{-\omega}$.
While the open circles are the Monte Carlo data, the solid lines are obtained
by using the results of the fit. Some data points are omitted for better visibility.}
\label{fig4}
\end{figure}

\begin{figure}
\begin{center}
\includegraphics[width=0.43\textwidth]{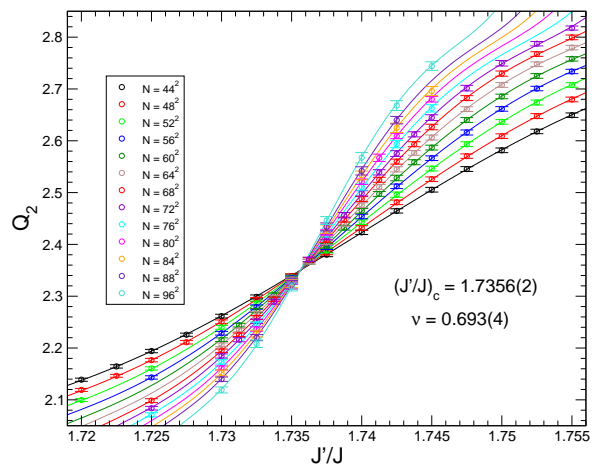}
\end{center}
\vskip-0.5cm
\caption{Fit of $Q_2$ to its finite-size scaling ansatz given by 
eq.~(\ref{FSS}) without taking the correction $cL^{-\omega}$ into account.
The number of spins $N$ used in the fit ranges from $44^2$ to $96^2$. 
While the open circles are the Monte Carlo
data, the solid lines are obtained by using the results of the fit.}
\label{fig5}
\end{figure}

\begin{figure}
\begin{center}
\includegraphics[width=0.43\textwidth]{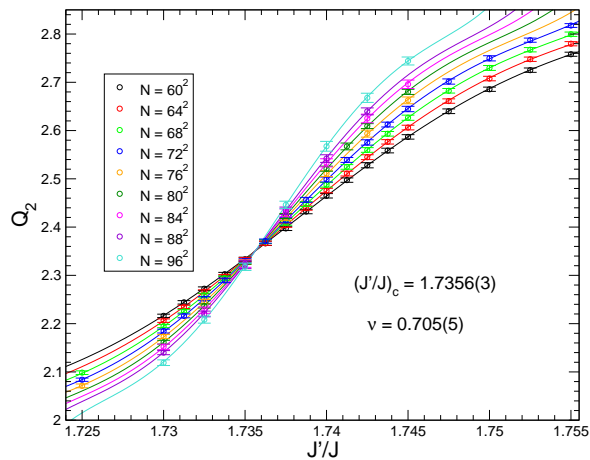}
\end{center}
\vskip-0.5cm
\caption{Fit of the numerical data for $Q_2$ to their finite-size scaling ansatz
without the correction term $cL^{-\omega}$ in eq.~(\ref{FSS}). The number of spins $N$ used in the fit 
ranges from $N=60^2$ to $N=96^2$.
While the open circles are the Monte Carlo data, the solid lines are obtained
by using the results of the fit.}
\label{fig6}
\end{figure}

\begin{figure}
\begin{center}
\includegraphics[width=0.43\textwidth]{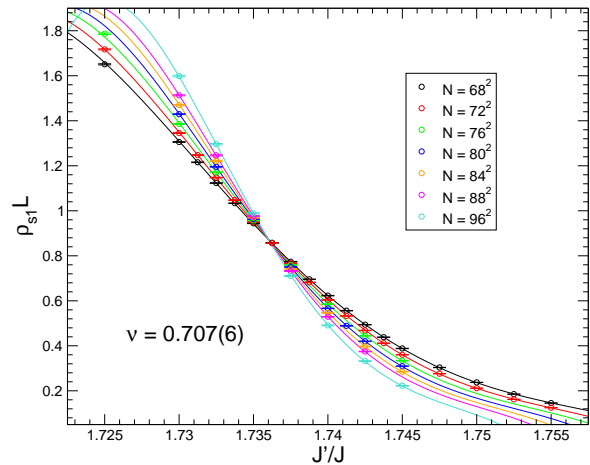}
\end{center}
\vskip-0.5cm
\caption{Fit of the numerical data for $\rho_{s1} L$ to their finite-size
scaling ansatz with the correction term $cL^{-\omega}$.
While the open circles are the Monte Carlo data, the solid lines are obtained
by using the results of the fit.}
\label{fig7}
\end{figure}

The simulations done in this work were performed using the loop
algorithm in ALPS library \cite{Troyer08}. 
In order to estimate where in the parameter space $J'/J$ the transition 
occurs, we investigate
the space-time volume dependence of the staggered susceptibility $\chi_s$.
To be more precise, while
$\chi_s$ grows with increasing space-time volume in the broken phase,
in the symmetric phase, $\chi_s$ should saturate for large
space-time volumes. We further scale $\beta$ with the 
square root of the number of spins $N$ in order to use this criterion
of space-time volume dependence of $\chi_s$. 
Using this criterion, we obtained figure \ref{fig1}. The trend
of the curves for different ratios $J'/J$ in figure 
\ref{fig1} indicates that
the transition takes place between $J'/J = 1.73$ and $J'/J = 1.74$.
Comparing this observation with the related critical point calculated
in \cite{Wenzel08} for the square lattice case, one sees that the honeycomb
lattice has a smaller value of $(J'/J)_c$. This is reasonable since the
antiferromagnet on the honeycomb lattice has a smaller value of the staggered 
magnetization density $\tilde{{\cal M}_s}$ per spin than that on the square lattice
\cite{Wie94,Jia08}. As a 
result, it is easier to destroy the antiferromagnism on the honeycomb 
lattice than on the square lattice by increasing the anisotropy $J'/J$. 
Therefore one expects a smaller critical value of $(J'/J)_c$ on 
the honeycomb lattice. 
Further, since phase transition only occurs for a system with infinite degrees of 
freedom, to investigate the critical behavior of a system 
on a finite volume, it is useful to employ the idea of finite-size 
scaling (FSS) hypothesis
for certain observables. Pioneering by Fisher in the seventies 
and generalizing later by Fisher, Barber and others including the 
derivation of FSS hypothesis from 
the framework of renormalization group
\cite{Fisher72,Brezin82,Barber83,Brezin85}, the FSS hypothesis
has become the essential
method to study the thermodynamic property of a finite system near the 
phase transition.   
The 
characteristic length for a finite system is the 
correlation length $\xi$. Consequently it is natural to consider its
thermodynamics property near the transition as a function of 
the variable $L/\xi$, where $L$ is the physical size of the system. 
The FSS hypothesis states that near the transition, any 
relevant property ${\cal O}_{L}(t)$
of a finite system with critical exponent $\kappa$ can be described by 
${\cal O}_{L}(t) = L^{\kappa/\nu} g_{\cal O}(L/\xi)$, where 
$t$ is given by $t = (j-j_c)/j_c$ with $j=J'/J$, $\nu$ is the corresponding
critical exponent for $\xi$,
and $g_{\cal O}$ is a
smooth function. Further, since $\xi$ diverges
as $\xi \sim |j-j_c|^{-\nu}$ near $j_c$, one arrives at
${\cal O}_{L}(t) = L^{\kappa/\nu} g_{\cal O}(Lt^{\nu}) 
= L^{\kappa/\nu} g'_{\cal O}(tL^{1/\nu})$, where $g'_{\cal O}$ is another
smooth function. FSS hypothesis can be rigorously derived by considering
the scaling of the relevant directions in the renormalization group flows. 
They can be understood intuitively as well by  
assuming the fluctuations of 
any relevant variable should be invariant at all length scales near $j_c$.
As a result, the appropriate variable for a finite system near $j_c$ 
is $tL^{1/\nu}$. 
Here we would like to emphasize that the function ${\cal O}_{L}$ appearing 
above depends on the lattice geometry as well as the boundary condition.
For example, for the Ising model on a 2-D periodic square lattice, 
there exists a so-called shift exponent 
$\lambda$ which is related to the deviation of the critical point on
a finite volume from the bulk value. For the observables
considered in this study, the most relevant FSS ansatz is given by
${\cal O}_{L}(t) = (1 + cL^{-\omega})g_{\cal O}(tL^{1/\nu})$, where
the confluent correction exponent $\omega$, which arises due to the inhomogeneous
part of the free energy as well as the nonlinearity of the scaling field,
is included explicitly. One should be aware that above expression of 
FSS ansatz is an asymptotical one which is valid only for large $L$ and close
to $j_c$. However to present the main results of this study, we find  
it is sufficient to employ the FSS ansatz introduced above for  
the data analysis. 
As mentioned earlier, when one approaches the
critical point $(J'/J)_c$ for a second order phase transition,
the observables $\rho_{s1} 3 N^{1/2}/2$ and $Q_2$ should satisfy the
following finite-size scaling formula 
\begin{eqnarray}
{\cal O}_{L}(t) &=& (1 + cL^{-\omega})g_{\cal O}(tL^{1/\nu}) \nonumber \\
&=& 
(1 + cL^{-\omega})\Big[g_0+tL^{1/\nu}g_1+(tL^{1/\nu})^2g_2\nonumber \\
&& + (tL^{1/\nu})^3g_3+\dots)\Big],
\label{FSS}
\end{eqnarray}
where ${\cal O}$ stands for either $\rho_{s1} 3N^{1/2}/2$ or 
$Q_2$, $t=(j-j_{c})/j_c$ with
$j = J'/J$ and $\nu$ and $\omega$ are the corresponding critical 
exponents introduced before. Further, 
in eq.~(\ref{FSS}),
$L = 3 N^{1/2}/2$ or $L = N^{1/2}$ depending on whether the observable ${\cal O}$
is referring to $\rho_{s1}3N^{1/2}/2$ or $Q_{2}$.  
Finally $g$ is a smooth function of the variable $tL^{\nu}$ and
$g_0+tL^{1/\nu}g_1+(tL^{1/\nu})^2g_2+(tL^{1/\nu})^3g_3+\dots$ 
is the Taylor expansion of the analytic function $g$. 
From eq.~(\ref{FSS}), one concludes that if the 
transition is second order, then the curves of different $L$
should intersect at the critical point for large
$L$. In the following, we will employ the finite-size scaling formula 
eq.~(\ref{FSS}) for various observables to calculate the critical point as 
well as the critical exponent $\nu$. 
After determining the regime where the transition occurs, we have 
performed substantial simulations with $ J'/J$ ranging from
$J'/J = 1.715$ to $J'/J = 1.755$ for $N = 10^2,~12^2,~14^2,\dots,~96^2$,
where $N$ is the total number of spins. We use sufficiently large $\beta$
so that all the observables measured in our simulations
take their zero-temperature values.
Figure \ref{fig2} shows the data points
for $\rho_{s1}L$ and $Q_2$ obtained from the simulations. 
From the figure, one observes that for both 
$\rho_{s1} L$ and $Q_2$ the curves of different $L$ indeed have a tendency to intersect at one point 
for large $L$, which in turn is an indications for a second order phase
transition. As a first step toward systematically analyzing the data,
we would like to have a better estimate of the critical point 
$(J'/J)_c$. This can be obtained by investigating the crossing points from
either $\rho_{s1} L$ or $Q_2$ at system sizes $L$ and $2L$. 
Figure \ref{fig3} shows 
such crossing points as functions of $1/L$. Indeed one sees both crossing
points from $\rho_{s1} L$ and $Q_2$ approach a common $J'/J$ with
increasing $L$. By using the analysis suggested in \cite{Wan05}, we find 
that $(J'/J)_c$ obtained from the upper (lower) crossing points in 
figure \ref{fig3} is within the range 
$(J'/J)_c \in [1.7345,1.7365]$ 
($(J'/J)_c \in [1.7355,1.7363]$) which agrees with the roughly
estimated $(J'/J)_c$ from $\chi_s$. Next, we turn to determine the 
critical exponent $\nu$ by employing the finite-size scaling ansatz for
the observables $\rho_{s1} L$ and $Q_2$. First of all, let us focus on
$Q_2 $. A fit of $Q_2$ to eq.~(\ref{FSS}) with $N \ge 12^2$ leads to
$(J'/J) = 1.7358(2)$ and $\nu=0.691(2)$.
The result is shown in figure \ref{fig4}.  
In the fits, we carefully make sure that
$\nu$ and $(J'/J)_c$ obtained from the fits are stable, namely
$\nu$ and $(J'/J)_c$ are consistent with respect to the elimination of
smaller $N$. 
For example, from the fit using data points of 
$N=24^2$ to $N=96^2$, we find $(J'/J)_c = 1.7357(2)$ and 
$\nu = 0.694(2)$. Both of them agree with the results obtained from the fit
using most of the available data points.
With $N \ge 26^2$, the fits are not stable. This is due
to the fact that for large $N$, the data points of $Q_2$ 
are not accurate enough to perform the fits including the term $cL^{-\omega}$  
in eq.~(\ref{FSS}), where the corrections become very small. 
Therefore we use another approach, namely we use the data 
with sufficiently large $N$ so that one can safely ignore the
correction $cL^{-\omega}$. We also make sure that the
results are stable like what we did before. The validity of this strategy
is justified as follows. We generate data points from the expression
$(1 + 0.5L^{-b})(1+(tL^{1/c})+0.1(tL^{1/c})^2+0.01(tL^{1/c})^3)$ for various
values of $L$ and $t$ with $b = 1.5$ and $c = 0.8$. By analyzing this set of
data, we do observe the convergence of $c$ to $c=0.8$ as one eliminates
more data points with smaller value of $L$.    
Surprisingly, with this strategy we find that for a wide range 
of $N$ we obtain consistent
results with our previous analysis including the correction term 
$cL^{-\omega}$.
For instance, using the data with $N \ge 44^2$, we find $\nu = 0.693(4)$
which agrees with $\nu=0.691(2)$ obtained earlier (figure \ref{fig5}). 
At this stage, one naturally would conclude that we find the same
unconventional critical behavior as that in \cite{Wenzel08}.
However, when we use larger $N$, we begin to observe that $\nu$ converges to 
the expected $O(3)$ value $\nu=0.7112(5)$. For example, fitting the data from
$N=60^2$ to 
$N=96^2$ to eq.~(\ref{FSS}) without the term
$cL^{-\omega}$ leads to $\nu=0.705(5)$, which is compatible with the
expected $O(3)$ value (figure \ref{fig6}). By applying the same analysis to the observable
$\rho_{s1}L$, we arrive at similar results. 
For example, although with a slightly large $\chi^2/{\text{d.o.f.}} \sim 5.1$,
by fixing $\nu=1.7356$ which is the expected $(J'/J)_c$ calculated from $Q_2$, 
we find $\nu = 0.694(4)$ for $N \ge 48^2$ from the fit including the 
term $cL^{-\omega}$ in eq.~(\ref{FSS}). We attribute the poor quality of
the fit to the correction to the critical point 
which are not considered here. Indeed with a fixed $(J'/J)_c = 1.7351$ which 
is not within the the statistical error of $(J'/J)_c$ obtained from $Q_2$, 
we arrive at $\nu=0.707(3)$ and a better $\chi^2/{\text{d.o.f.}} = 3.7$. 
Notice this $\nu$ is even compatible with the known $O(3)$ value.
With a fixed $(J'/J)_c = 1.7356$,
only from $N \ge 64^2$ we begin to obtain $\nu$
consistent with the expected $O(3)$ value $\nu = 0.7112(5)$ : by fitting data
points with $N \ge 68^2$ to eq.~(\ref{FSS}) with the
term $cL^{-\omega}$, we obtain $\nu=0.707(6)$ (figure \ref{fig7}).

Table 1 summarizes the results of all the fits mentioned earlier.
We would like to point out that the statistical errors shown in this study
are obtained by binning. Hence the error bars presented here might be less
accurately determined compared to those calculated with more
sophisticated methods. 
Indeed from the $\chi^2/{\text{d.o.f.}}$ shown in table 1, 
one sees that while the errors for $Q_2$ seem to be overestimated, the 
errors for $\rho_{s1}L$ are likely underestimated. Nevertheless, 
from the $\chi^2/{\text{d.o.f.}}$ presented in table 1, the results we have 
concluded should remain valid even an optimization procedure is applied for 
the data analysis.
Notice the exponent $\omega$ for the last 2 rows in table 1  
are not shown explicitly since
the corresponding error bars are of the same magnitude 
as $\omega$ themselves. 
This is likely due to the fact that the data is not precise enough to
determine $\omega$ unambiguously. Further, the $\omega$ we obtain from the
fits are not consistent with the theoretical value. For example, the $\omega$ 
in the last row of table 1 is $0.38$ which is significantly smaller than the 
theoretical value $\omega \sim 0.78$. However since 
the value of $\omega$ will be affected by higher order operators to some 
extent as well as there might be a strong correlation between $c$ and $\omega$
in formula~(\ref{FSS}), 
the $\omega$ presented here should be treated as a effective one. 
Actually we find that both $\chi^2/{\text{d.o.f.}}$ and $\nu$ from
the fit of the last row in table 1 are very stable with a fixed value of
$\omega$ including $\omega = 0.78$. Therefore the result of $\nu = 0.707(6)$ 
shown in table 1 should be trustable.

\begin{table}
\begin{center}
\begin{tabular}{|c|c|c|c|c|c|c|}
\hline
$ {\text{Observable}}$ & $N_{\text{min}}$ & $N_{\text{max}}$ & $(J'/J)_c$ & $\nu$
& $\omega$ & $\chi^2/{\text{d.o.f.}}$\\
\hline
\hline
$Q_2$  & $12^2$ & $96^2$ & $1.7358(2)$  & $0.691(2)$  & 1.98(5) & 0.12  \\
\hline
$Q_2$  & $24^2$ & $96^2$ & $1.7357(2)$  & $0.694(2)$  & 3.5(6) &  0.12 \\
\hline
$Q_2$  & $44^2$ & $96^2$ & $1.7356(2)$  & $0.693(4)$  & - & 0.1 \\
\hline
$Q_2$  & $60^2$ & $96^2$ & $1.7356(3)$  & $0.705(5)$  & - & 0.08 \\
\hline
$\rho_{s1}L$  & $48^2$ & $96^2$ & $1.7356\,({\text{fixed}})$  & $0.694(4)$  & 0.9(3) & 5.1 \\
\hline
$\rho_{s1}L$  & $48^2$ & $96^2$ & $1.7351\,({\text{fixed}})$  & $0.707(3)$  & - & 3.7 \\
\hline
$\rho_{s1}L$  & $68^2$ & $96^2$ & $1.7356\,({\text{fixed}})$  & $0.707(6)$  & - & 1.8 \\
\hline
\end{tabular}
\end{center}
\label{table1}
\caption{Results of the fits mentioned in the text. $N_{\text{min}}$
and $N_{\text{max}}$
refer to the smallest and largest number of spins in the fit, respectively.}
\end{table}

\section{Discussion and Conclusion}
\label{discussion}

In this note, we have performed large scale Monte Carlo simulations to study
the critical behavior of the spin-1/2 Heisenberg model with a spatially
staggered anisotropy on the honeycomb lattice. In particular, we
have demonstrated the subtleties of determining the critical exponent $\nu$
from investigating the finite-size scaling behavior of the relevant 
observables. From what we have found, we conclude that for the second order
phase transition considered here, the large volume data
is essential to calculate the corresponding critical exponent $\nu$ 
correctly. With a detailed 
data analysis, we find that the critical exponent $\nu$ is given by
$\nu=0.707(6)$ which is cosistent with the most accurate $O(3)$ value 
$\nu=0.7112(5)$ obtained in \cite{Cam02}. 
By including the sub-leading correction in the finite-size
scaling ansatz and using most of the available data points, we can obtain a 
value of $\nu$ consistent with that calculated in \cite{Wenzel08}. 
Concerning the finite volume effect, we believe the $\nu$ obtained from 
using only large $N$ is more reliable. 
Our results do not necessarily imply that 
the critical exponent $\nu$ obtained in \cite{Wenzel08} should not be
trusted. 
However from what we have found in this study,
it would be desirable to go beyond the volumes used in \cite{Wenzel08}
to see whether the unconventional behavior observed in \cite{Wenzel08} will
continue to hold with the addition of larger volumes or not.

\section{acknowledgments}
We like to thank U.-J. Wiese for useful discussions. 
This work is supported in part by funds provided by the Schweizerischer 
Nationalfonds (SNF). 
The ``Center for Research and Education in Fundamental
Physics'' at Bern University is supported by the ``Innovations- und 
Kooperationsprojekt C-13'' of the Schweizerische Universit\"atskonferenz 
(SUK/CRUS).

\end{document}